# A Comprehensive Database of Leaf Temperature, Water, and $CO_2$ Fluxes in Young Oil Palm Plants Across Diverse Climate Scenarios

Raphael Perez[1,2], Valentin Torrelli[1,2,3,4], Sandrine Roques[1,2], Sébastien Devidal[5], Clément Piel[5], Damien Landais[5], Merlin Ramel[3,4], Thomas Arsouze[3,4], Julien Lamour[6], Jean-Pierre Caliman[7], Rémi Vezy[3,4]

[1]CIRAD, UMR AGAP Institut, F-34398 Montpellier, France
[2]UMR AGAP Institut, Univ Montpellier, CIRAD, INRAE, Institut Agro, F-34398 Montpellier, France
[3]CIRAD, UMR AMAP, F-34398 Montpellier, France 4
[4]AMAP, Univ. Montpellier, CIRAD, CNRS, INRAE, IRD, F-34398 Montpellier, France
[5]Ecotron Européen de Montpellier, Unité Propre de Service 3248, Centre National de la Recherche Scientifique (CNRS), Campus Baillarguet, F-34980 Montferrier-sur-Lez, France;
[6] Centre de Recherche sur la Biodiversité et l'Environnement (CRBE), Université de Toulouse, CNRS, IRD, Toulouse INP, Université Toulouse 3 – Paul Sabatier (UT3), Toulouse, France.
[7]SMART Research Institute, Pekanbaru 28112, Indonesia

## Abstract

Functional-structural plant models (FSPM) replicate plants' responses to their environment and are useful for predicting behavior in a changing climate. However, they rely on detailed measurements of traits, which are difficult to collect consistently across scales, often limiting model parameterization and thorough evaluation, and thereby reducing confidence in model predictions.

Here, we provided a comprehensive dataset allowing to generate a digital twin of an experiment conducted on four oil palm plants (*Elaeis guinnensis*) grown under multiple controlled environmental scenarios, including varying $CO_2$ concentrations, light, temperature and humidity conditions. The dataset included detailed reconstructions of the three-dimensional plant structures derived from terrestrial LiDAR point clouds, and enabled the parametrization of biophysical processes at the leaf scale such as photosynthesis and stomatal conductance, as well as the collection of plant-scale measurements (gas exchange measurements of $CO_2$ and $H_2O$), which can be compared with FSPM simulations.

The tree-dimensional reconstructions effectively represented the architecture of the plants and showed strong correlation with the measured total leaf area. Flux measurments allowed to quantify the impact of climate variables (temperature, vapor pressure deficit, radiation) on plant transpiration and photosynsthesis. Such data could be used for future comparisons between simulated and observed physiological behaviours could be used to evaluate the quality of the physiological formalisms independently. By bridging the scales from individual leaves to the entire plant, this database allows modellers to both calibrate their biophysical models at a fine spatial resolution and evaluate their predictive accuracy at the plant scale. The provided data will facilitate benchmarking of biophysical models, help identify sources of model uncertainty, and ultimately enhance model predictions, which can be applied in various fields, from cognitive studies to decision support applications.

## Introduction

Biophysical plant models aim to mechanistically represent how plants acquire, process, and utilise biophysical resources—such as light, water, and carbon—across spatial and temporal scales by integrating fundamental physiological processes such as photosynthesis, transpiration or energy balance (Fourcaud et al., 2008). These models are used by different communities of scientists interested in the simulation of plants at different scales: organ-to-plant with functional-structural plant models (FSPM; Vos et al., 2010), plant-to-plot with individual process-based models (PBM; *e.g.* Duursma & Medlyn, 2012, Maréchaux & Chave, 2017), or earth models (*e.g.* Krinner et al., 2005). Generally, FSPM and PBM use leaf-scale measurements to parameterise different sub-models, enabling the upscaling of such measurements to the plant (FSPM) or plot level (PBM), thereby simulating variables that are hard or impossible to directly measure (*e.g.* water and energy balance) and predicting system behaviour under current or new conditions (*e.g.* assessing climate change impacts) (Wu, 2023). The community of plant growth modeling is encountering challenges in parameterizing models due to the complexity of these models and the high costs associated with data acquisition. Consequently, there is a notable absence of benchmarking, which hinders the ability to compare the diverse range of models present in the literature effectively (Cournède et al., 2013).

In FSPM, evaluating a model becomes challenging when numerous interconnected processes are simulated, as it is often the case with biophysical processes in natural systems, and these interactions may act across different scales and environments (Rötter et al., 2012). Yet, despite their broad relevance, these models are often evaluated using data collected at a single scale—most commonly at the leaf or plot level—due to the rarity of datasets that capture both detailed organ-level measurements and integrated responses at the whole-plant scale (Wu, 2023). In practice, the challenge of acquiring coherent datasets that simultaneously capture plant geometry, physiological traits, and whole-plant gas exchange under controlled and well-documented conditions often restricts the thorough assessment of models. There is no standard method to test FSPMs since collecting all the necessary data for model development is often challenging or impossible, leading to many assumptions and significant uncertainty in the resulting models (Wang et al., 2018). Consequently, most models remain untested or insufficiently evaluated at integrative levels, reducing our confidence in their predictions and their applicability to real-world scenarios.

The strength of FSPM lies in their ability to simulate explicit plant morphogenesis and the development of plant architecture (Soualiou et al., 2021). When combined with biophysical processes, such models can effectively capture plant functioning in specific agronomic designs where the assumptions of conventional crop models may not be applicable. For example, canopy photosynthesis modeled through empirical light extinction and leaf area index cannot be effectively applied in complex systems characterized by heterogeneous canopies, such as agroforestry systems (Vezy et al., 2023). The recent development of diversified systems, such as those derived from intercropping and agroforestry, presents challenges for the plant modeling community (Enders et al., 2023). Modeling these systems would thus require an explicit representation of the spatial variability of resources to accurately account for competition among plants.

Another limitation of current holistic modeling approaches is the lack of available datasets for testing extreme climatic events. Since physiological processes typically exhibit nonlinear responses to climatic variables, it is crucial to acquire this data and assess model performance in predicting plant behavior under future climate scenarios.

We argue that accessible databases that enable the evaluation of biophysical models across various scales - from the leaf level to the plant and plot levels - are crucial for increasing confidence in model predictions. Those databases should include different experiments conducted in more to less controlled conditions, allowing the evaluation of models with more or less degrees of freedom, evaluating physics-based processes first, and coming to more biology-based processes. For example, the experimental data acquired by Schymanski and Or, (2017) can help evaluate the energy balance (sensible and latent heat) components of models at the scale of an individual leaf, thanks to their experiment on highly controlled conditions using an artificial leaf.

In this paper, we address the following critical gap by providing a comprehensive database of biophysical measurements in young oil palm plants (*Elaeis guineensis*) to evaluate biophysical processes at leaf-to-plant scale under controlled conditions. Our dataset encompasses detailed 3D reconstructions of plant structure, leaf-level gas exchange measurements that inform fundamental physiological parameters, and concurrent whole-plant flux data acquired under controlled dynamically varying climatic conditions. By bridging the scales from individual leaves to the entire plant, this database allows modellers to calibrate a digital twin of the experiment, allowing the assessment of their biophysical models at a fine spatial resolution and evaluate their predictive accuracy at a more holistic level: the whole plant. In doing so, we take a crucial step toward a new generation of open-access databases that empower researchers to rigorously benchmark biophysical models, ultimately improving their robustness, reliability, and utility.

## Data for calibrating a digital twin of the experiment

### Plant material and pre-experiment growing conditions

Four oil palm plants (*Elaeis guineensis Jacq.*) from two genetic origins were studied: Deli x Lame crossing (P1, P2 & P4) and a Deli x Yamgambi crossing (P3). Only four plants were selected due to the time available for carrying out the experiment. The individuals were selected based on their visually distinct morphology, which stood out among all the available plants. We did not intent to have replicates in this experiment, but rather explore the potential diversity in plants responses as a consequence of their structural and physiological differences. The plants were sown on May 11$^{th}$ 2020 and cultivated in a greenhouse from CIRAD's Abiophen platform (Montpellier, France) with a 12 hours photoperiod (light from 6 am to 6 pm). The air temperature was controlled at 26°C during the day and 21°C during the night, relative humidity at 70%, photosynthetically active radiation (PAR) at 600 µmol m$^{-2}$ s$^{-1}$, and air $CO_2$ concentration ([$CO_2$]) at around 400 ppm. The plants were irrigated every two to three days to prevent water stress. On February 25th 2021, the plants were transferred to the microcosm's experimental platform of the European Ecotron of Montpellier (https://www.ecotron.cnrs.fr).

### Microcosms

#### *Set-up*

Two microcosms growth chambers of dimensions 114 cm width x 113 cm depth x 152 cm (~1.5 m$^3$) height were used for the two-month experiment. The microcosms allowed for a precise control of radiation in the visible spectrum (Supplementary Material Figure S1) with four LED lamps (Soledaire, France), air temperature (5-50±0.5°C), relative humidity (20-90±3%), and $CO_2$ concentration (10-2000ppm). The monitoring microcosm was used to measure the biophysical processes of a single plant in response to different climate conditions with varying air temperature, relative humidity and radiation. The storage microcosm was used to store the three other plants waiting for their turn in the monitoring microcosm.

#### *Monitoring*

The monitoring microcosm was operated as an open $CO_2$ gas exchange system. The flow rate of dry air at the inlet was measured and regulated at 4.9 Nm$^3$ h$^{-1}$ using a mass flow regulator (F-202AV, Bronkhorst, The Netherlands). The net $CO_2$ flux was measured continuously by sequentially measuring the inlet and outlet of the chamber every 5 minutes using a Valco selector (EUTA-SD4MWE, VICI, Switzerland) and a Picarro G2101-i (Picarro, USA) $CO_2$ analyser. For each position of the selector, the first two minutes were discarded, and the last three minutes were averaged. The air sampling at the inlet circulated first through a 30-litre buffer volume with a flow rate regulated at 1.5 l min$^{-1}$ using a needle valve, while the outlet was directly measured.

The monitoring microcosm was also equipped with a photosynthetically active radiation sensor (quantum sensor LICOR LI-190 SA, Lincoln, NE, Figure 1), air temperature (CTN 35, Carel) and humidity sensors (PFmini72, Michell instrument, USA), a thermal camera on the top left corner pointing towards the centre of the chamber to measure leaf temperature, and a precision scale to monitor weight. The thermal camera and the precision scale were controlled by a Raspberry Pi (https://www.raspberrypi.org/) board that triggered a camera shot every minute and automatically logged the stream of data from the scale. Data from all other sensors were automatically logged by the microcosm facility.

The pot of the plant was sealed before entrance into the monitoring microcosm to avoid water loss to the atmosphere, enabling the computation of plant transpiration from weight loss. Plants were automatically watered every six hours to maintain non-limiting soil water availability.

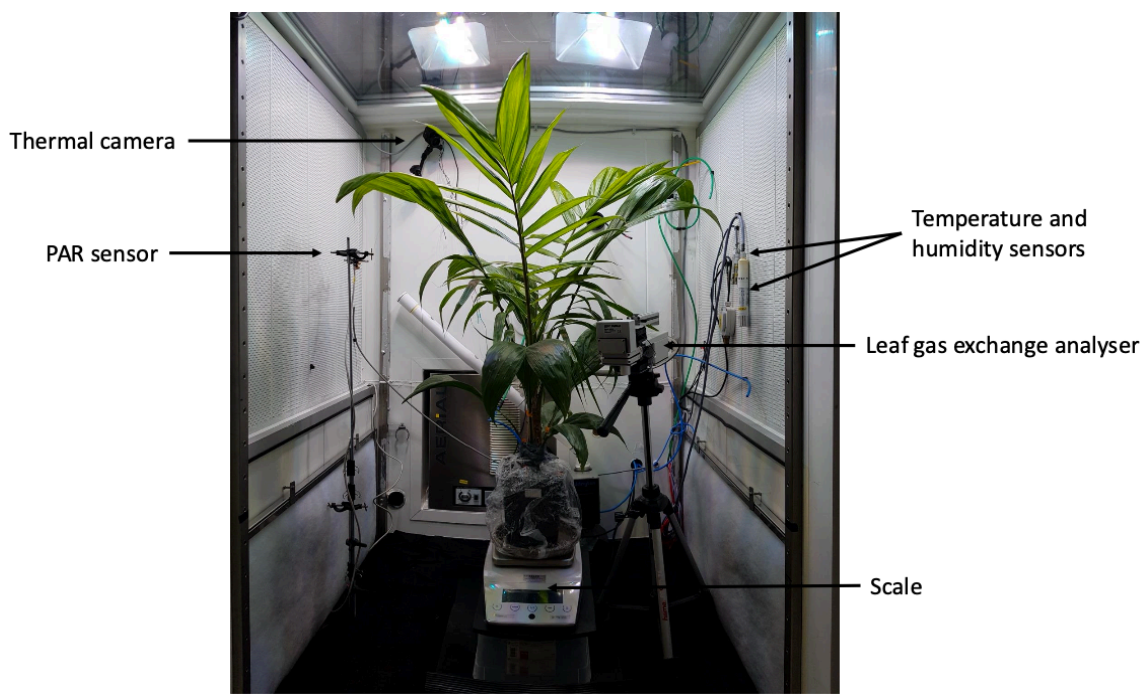

*Figure 1: Oil palm plant in the monitoring microcosm. The pot was sealed to avoid water loss to the atmosphere. A precision scale was positioned under the pot to estimate plant transpiration from variations in plant weight. Sensors for photosynthetically active radiation (**LICOR** LI-190), temperature, and relative humidity were installed in the chamber to regulate the environmental conditions. The head of the leaf gas exchange analyser (Walz GFS-3000 ) was positioned in the chamber to conduct either $CO_2$ response curves in the storage microcosm or to follow leaf assimilation during specific scenarios in the monitoring chamber.*

## Climate scenarios

The microcosm climate conditions were established based on the average daily variations recorded at a weather station in Pekanbaru, Indonesia, a region characterized by optimal conditions for oil palm cultivation. The baseline environmental parameters comprised a constant atmospheric $CO_2$ concentration of 400 ppm, daily air temperature fluctuations ranging from 22°C to 33°C, relative humidity varying between 82% and 51%, and PAR levels ranging from 0 to approximately 300 µmol m⁻² s⁻¹ measured at mid-height within the chamber, and approximately 1000 µmol m⁻² s⁻¹ directly beneath the light source. These baseline conditions were maintained consistently within the storage microcosm throughout the entire experimental period.

To investigate the effects of key climate variables on physiological processes such as photosynthesis and stomatal conductance, the baseline conditions were systematically modified to simulate a range of environmental scenarios. Specifically, atmospheric $CO_2$ concentrations were elevated to 600 ppm and 800 ppm to assess the influence of increased $CO_2$ availability. Radiation intensity was reduced to simulate cloudy conditions, with PAR set to 130 µmol m⁻² s⁻¹ at mid-day and mid-height within the chamber. Temperature variations were introduced by decreasing or increasing the baseline air temperature by 30%, representing colder and hotter conditions, respectively. Additionally, combined adjustments of temperature and relative humidity were applied to simulate drier and cooler environments by increasing relative humidity by 30% while decreasing temperature by 30%, as well as drier and hotter conditions by decreasing relative humidity by 30% and increasing temperature by 30%. These scenarios were designed to capture the complex interactions between $CO_2$ concentration, radiation, temperature, and humidity, with particular emphasis on VPD, given the pronounced sensitivity of oil palm stomata to leaf-to-air vapor pressure differences (Dufrêne & Saugier, 1993). The eight resulting climate scenarios are presented Figure 2.

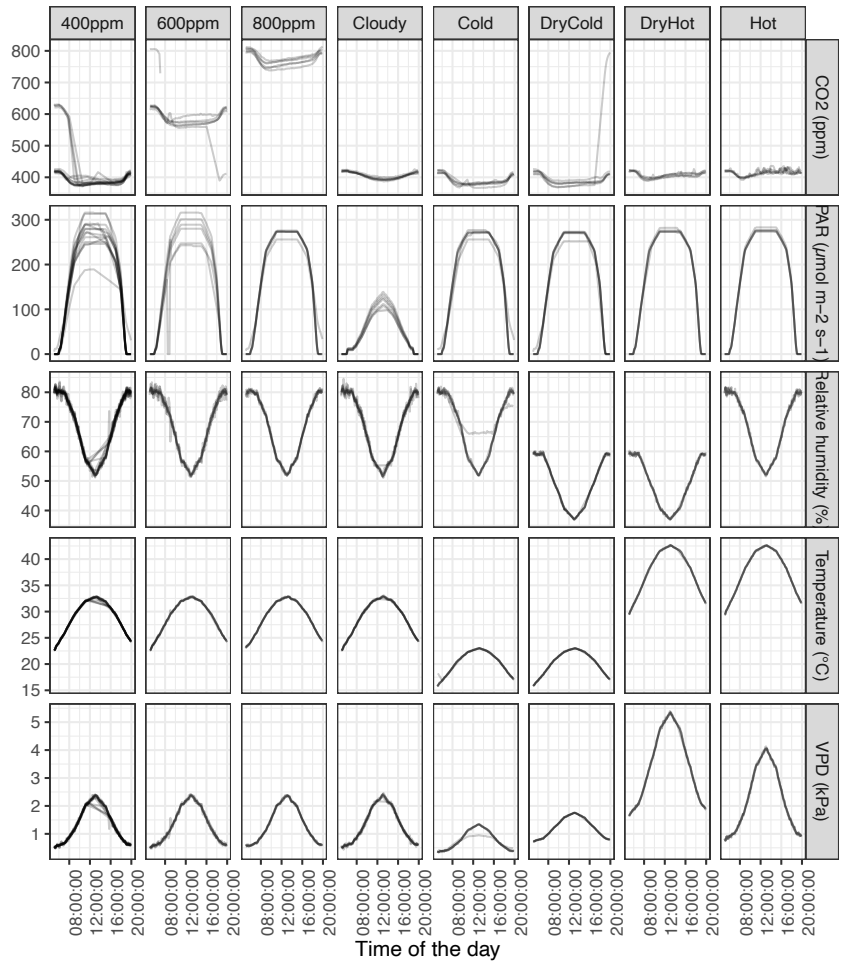

*Figure 2: Monitoring radiation, air temperature, and relative humidity over time for the eight climate scenarios. Each line represents a day of measurement. Photosynthetically active radiation was measured at the chamber's centre height. The reference scenario is the '400ppm' scenario (first column). The vapor pressure deficit (VPD) was calculated as a function of air temperature and air relative humidity.*

Each plant was sequentially placed in the monitoring microcosm for one or more days, following the experimental design illustrated in Figure 3. Scenarios involving potentially detrimental effects on plant function due to extreme high temperatures were conducted during the final days of measurements for each plant. Additionally, some scenarios were repeated for individual plants to assess temporal changes in physiological responses over the course of the experiment.

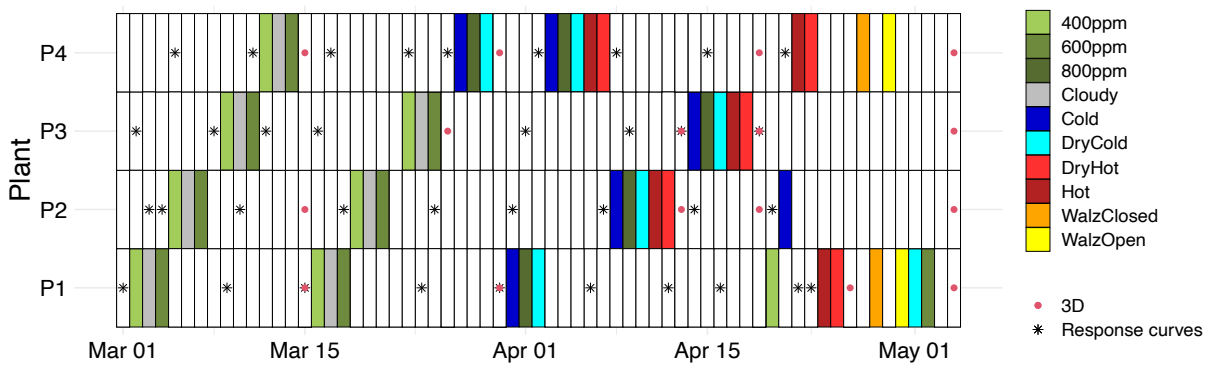

*Figure 3: Climate scenarios set in the monitoring microcosm and the sequence of measurements. Grey cells indicate dates on which the plant is in the storage microcosm. Points indicate the date of leaf gas exchange measurement (black crosses) and dates of 3D reconstruction of plants (red points). Colors refer to the climate scenarios (for WalzClosed and WalzOpen scenarios, refers to section Leaf-level CO$_2$ and H$_2$O gas exchanges).*

## Leaf-level CO$_2$ and H$_2$O gas exchanges

Leaf gas exchange measurements were performed with a Walz GFS-3000 portable gas analyser with a Walz PAM-Fluorometer 3056-FL (Walz, Effeltrich, Germany) and a cuvette area of 8 cm$^2$. One leaf per plant was measured in the storage microcosm during the experiment (responses curves, Figure 4) before and after the plant was placed under a climate scenario sequence. At each date, the last leaf fully expanded was selected for conducting photosynthesis response to CO$_2$ (A~C$_i$ curves), followed by the photosynthesis response to photosynthetic photon flux density (A~PPFD curves) and stomatal conductance response to vapour pressure deficit (G$_s$~VPD).

The A~C$_i$ curves were performed at a saturating PPFD of 1500 µmol m$^{-2}$ s$^{-1}$, a controlled cuvette air temperature of 25°C, a relative humidity of 65%, a constant air flow rate through the cuvette of 750 mL min$^{-1}$, and changing [CO$_2$] from 400 to 50 ppm, then from 400 to 2000 ppm in 13 steps of [CO$_2$] in total. Measurements during the A~C$_i$ curves were performed every 90 seconds.

The A~PPFD curves were performed after an acclimation to ambient CO$_2$ of 6 minutes after the A-Ci curves. The temperature and relative humidity were maintained at 25°C and 65% as for the A-Ci curves and the light was changed in 9 steps of PPFD from 1500 µmol m² s$^{-1}$ to 10 µmol m² s$^{-1}$. Measurements during the A~PPFD curves were performed every 180 seconds.

The G$_s$~VPD curves were measured in 7 steps from 0.7 kPa to 2.5 kPa at 1500 µmol m² s$^{-1}$ of PPFD and 400 ppm [CO$_2$]. The VPD was controlled by changing the relative humidity (from 75% to 30%) and the air temperature (from 23°C to 27°C). Measurements during the G$_s$~VPD curves were performed every 120 seconds.

The three response curves can be used to estimate the parameters of coupled leaf photosynthesis and transpiration models (Busch et al., 2024). In the dataset, we used the A~C$_i$ curves to estimate the Farquhar-von Caemmerer-Berry (FvCB) of C3 photosynthesis (Farquhar et al., 1980) parameters at a reference temperature of 25°C using the temperature-dependent parameters from Kumarathunge et al., 2019), except for the rate of decrease of the function above the optimum for the rate of electron transport (Hd$_j$) and rubisco activity (Hd$_v$) that were taken from (Dreyer et al., 2001) and (Medlyn et al., 2002). The estimated parameters included the maximum rate of RuBisCO carboxylation ($V_{cmax}$), the maximum potential electron transport rate ($J_{max}$), the rate of mitochondrial respiration ($R_d$) and the triose phosphate utilisation rate (*TPU*, Figure 4a). Response curves to VPD were used to estimate the parameters of Medlyn's stomatal conductance model (Medlyn et al., 2011), *i.e.* the residual stomatal conductance ($g_0$) and the slope parameter ($g_1$, Figure 4b), although other models could be used.

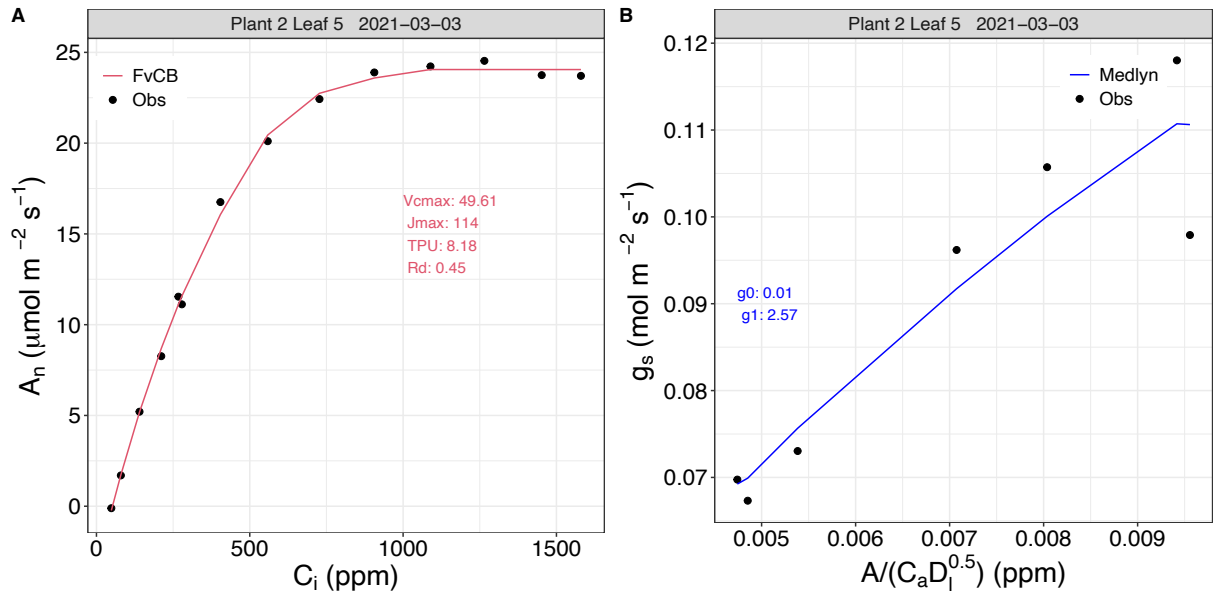

*Figure 4: Calibration of photosynthesis and stomatal conductance models from leaf gas exchange responses curves. A) A~$C_i$ response curve fitted with the Farquhar-von Caemmerer-Berry (FvCB) model. B) The $G_s$~VPD (Dl) response curve is fitted with Medlyn's stomatal conductance model (Medlyn et al., 2011).*

At the end of the experiment, from 27th to 30th April, additional measurements were conducted on two plants (Figure 3) to assess the correlations between leaf-level gas exchanges and plant-level gas exchanges. The aim of these measurements was to investigate whether leaf gas exchange is influenced by overall plant conditions, mainly focusing on the light environment. The plant was placed in the flux chamber, with one leaf attached to the Walz leaf gas analyser. At the leaf scale, the conditions within the Walz's head remained constant in terms of temperature, [$CO_2$] and relative humidity, while the light was either at saturation (1500 μmol m$^{-2}$ s$^{-1}$; WalzClosed) or following the light conditions in the microcosm by removing the light component from the head (WalzOpen). At the plant scale, the climate conditions (temperature, relative humidity, [$CO_2$]) within the microcosm followed the reference scenario (400 ppm), except for the radiation that drastically change from shading periods to full light periods (Figure S2).

## Leaf chlorophyll content

The chlorophyll content of leaves was measured with a SPAD chlorophyll meter (SPAD-502; Minolta, Ltd., Japan). At the beginning of the experiment (February 16th and 23rd), SPAD readings were taken on every leaf of all the plants. The SPAD value was calculated as the average of four measurements taken from the middle section of the leaf, corresponding to the area where gas exchange measurements were conducted. Then, SPAD measurements were repeated on all leaves of each plant prior to conducting leaf gas exchange measurements in the microcosm. These data were collected to evaluate formalisms that enable the upscaling of leaf gas exchange from the leaf level to the plant scale, specifically examining the relationship between variations in photosynthetic parameters and chlorophyll content (Prieto et al., 2012)

## Plants architecture

Three-dimensional reconstructions of the plants were essential for simulating biophysical processes such as light interception and photosynthesis. To capture the evolution of plant architecture over time, LiDAR scans of the four plants were conducted weekly during the entire period of flux measurements. At least three viewpoints were captured for the co-registration process to accurately represent the whole plant and minimise occlusion issues. At the end of the experiment, the leaves were removed from the plant and scanned individually, a step that enabled detailed reconstruction of each leaf for each date—particularly in the densely overlapping central regions— and the bulb without interference from adjacent foliage.

Plant reconstructions were carried out manually in Blender (Blender Development Team, 2022). Using plane meshes fitted to the leaf point clouds via the *poly build tool* with automatic vertex merging (Figure 5A), each organ was reconstructed separately and exported as a `ply` file.

To overcome the challenge of distinguishing overlapping leaves in the central region, we leveraged the individual leaf reconstructions to guide the plant-scale point clouds reconstructions. Starting with the latest LiDAR scan (which was closest in time to the individual leaf scans), we integrated these detailed leaf models into the overall reconstruction. Then, proceeding chronologically backward, we manually modified and adjusted the meshes to fit the point clouds of preceding dates, using the later reconstructions as references. This sequential, reference-based approach enhanced the consistency and accuracy of the reconstructions over time (Figure 5B). This was possible because the morphology of oil palm leaves does not elongate and expand over time once they are emitted.

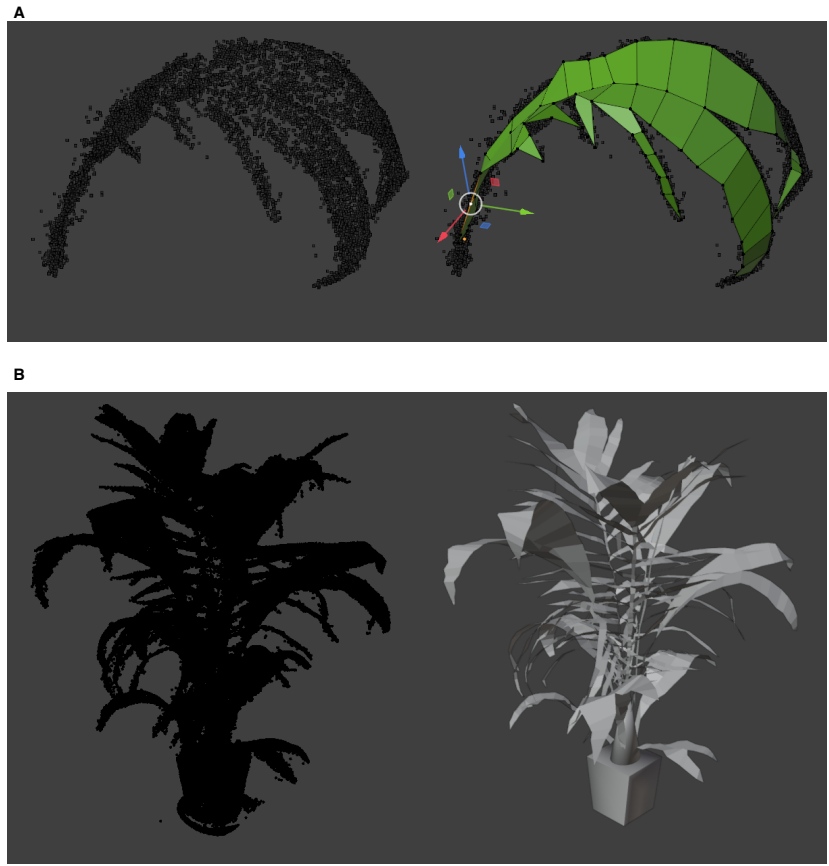

*Figure 5: 3D reconstruction from LiDAR point clouds. A) Building plane meshes on point cloud with the poly build tool of Blender. B) Full reconstruction of the 3D mock-up from points cloud.*

Each time a plant was placed in the microcosm to undergo a climate scenario sequence, we selected the LiDAR point cloud that best represented these dates for reconstruction. Due to the slow development of the plants, four dates were chosen to capture the evolution of the plant architecture for each plant (Figure 3).

In the final step, each ply reconstruction file was converted into an Open Plant Format (OPF, Griffon & de Coligny, 2014), a portable file that stores both plant topology and geometry and is commonly used in simulation models of biophysical processes. The plant topology was defined by six symbols: Plant, Pot, Bulb, Stipe, Leaf and Spear (leaf without expanded leaflets).

The quality of the virtual reconstructions was assessed by comparing the area of each virtual leaf to measurements obtained with a leaf area meter (Licor LI-3100C) at the end of the experiment. The results demonstrated a high level of agreement between the measured and reconstructed leaf areas ($R^2$=0.99 for the four plants; Figure 6).

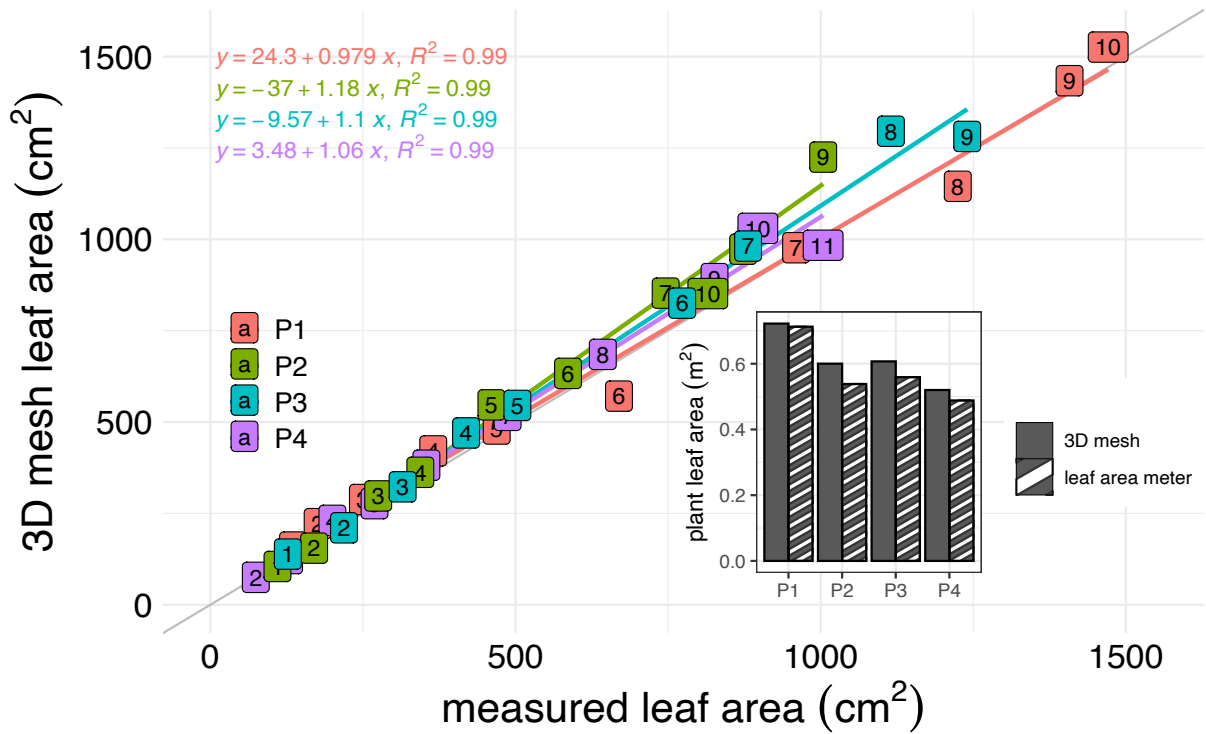

*Figure 6: Evaluation of 3D reconstructions based on leaf area. Leaf numbers are indicated within each data point, while colors correspond to individual plants. The inset compares the total leaf area per plant estimated from the 3D mesh with measurements obtained using a leaf area meter at the end of the experiment (4th May).*

## Mapping photosynthetically active radiation in the microcosm

The computation of the digital twin necessitated a precise reproduction of the conditions experienced by the plants. In addition to the climate inputs discussed in the previous sections, it was essential to accurately simulate the radiative environment to reliably estimate the light intercepted by the plants and the subsequent process of photosysnthesis. We conducted specific measurements to assess the spatial heterogeneity of light within the microcosm using a ceptometer (Sunscan, Delta-T, Figure 7). This device was equipped with 60 PAR sensors, each spaced 1.6 cm apart, which allowed for a fine-scale mapping of the light environment. These measurements were performed to thoroughly evaluate 3D reconstruction and their impact on light interception and light transmission within the microcosm. The idea was to provide values of light intensity at multiple points in the space, with or without plant in the microcosm, that could be compare with light intensity simulated from 3D simulations.

Measurements were performed under three distinct conditions: i) in an empty microcosm to capture both direct and diffuse radiation; ii) in an empty microcosm with black felt applied to the walls to suppress scattered light and only measure direct light (Figure 7A); and iii) in the microcosm with a plant inside it (Figure 7B).

For the empty chamber, light was measured at four vertical heights (21 cm, 51 cm, 81 cm, and 111 cm from the light source) to capture the vertical distribution of radiation. In the presence of a plant, measurements were taken at the top of the pot (105.4 cm from the light source), and at mid-canopy level (91.5 cm from the light source). At each of these heights, we conducted a horizontal mapping by measuring light at eleven positions spanning from 10 cm to 90 cm across the chamber starting from the left side, with additional measurements at 5 cm and 95 cm to capture edge effects. The pot was placed in the centre of the chamber, approximately 56 cm from the lateral and back walls, and a black net was positioned on the chamber floor in all conditions to minimize light reflections. The light source consisted of four LED spots, selected for its spectral distribution close to natural light (Supplementary Material Figure S1).

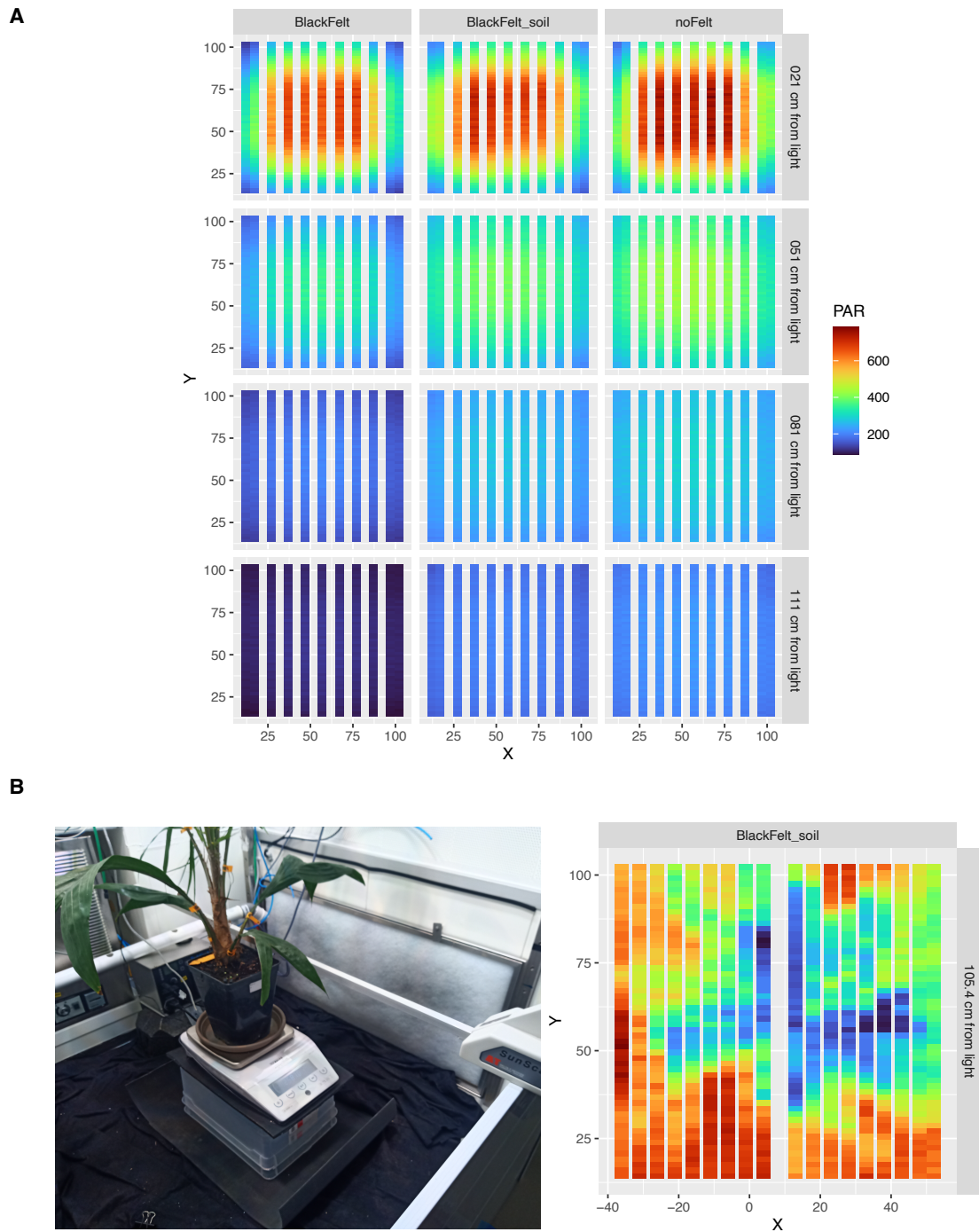

*Figure 7: Mapping of light distribution within the microcosm using a SunScan system (Delta-T) equipped with 60 light sensors. (A) Radiation maps illustrating the spatial variation of light intensity relative to the distance from the light source and the optical properties of the chamber walls and floor, which were either uncovered or covered with black felt to minimize light diffusion (blackfelt: walls and fllor covered, blackFelt_soil: only floor covered; noFelt: uncovered walls and floor). (B) Left panel: measurements of light transmission beneath the plant P1 obtained with the SunScan; right panel: corresponding spatial map of photosynthetically active radiation (PAR) intensity beneath the plant. Colors indicate PAR levels in µmol m$^{-2}$ s$^{-1}$.*

## Data for evaluating biophysical processes from leaf to plant scale

## Leaf temperature

Leaf temperature primarily varies with air temperature, but it is also influenced by the state of the stomata, as stomatal opening can cool the leaf through transpiration. Typically, leaf temperature is an output of biophysical models. Therefore, measuring leaf temperature over time and assessing the spatial distribution of these temperatures within the plant can serve as valuable variables for evaluating the accuracy of biophysical processes.

Leaf temperature was measured with a FLIR Vue™ Pro R thermal camera triggered by a Raspberry Pi to take one image every second automatically. The camera was installed on the top left corner of the chamber and oriented toward the centre of the microcosm to ensure optimal coverage of the plant canopy. A calibration process was performed using objects with known temperature. Images were continuously recorded from March 2$^{nd}$ to May 3$^{rd}$ and later processed to extract regions corresponding to identifiable leaves from each frame. For every image, fixed masks were defined manually to isolate the maximum consistently visible area of each leaf over time, accounting for slight movements induced by wind inside the chamber (Figure 8). As a result, leaf temperatures were calculated over time for each pixel of the masks, after adjusting for air temperature and relative humidity within the chamber. Finally, the mean,

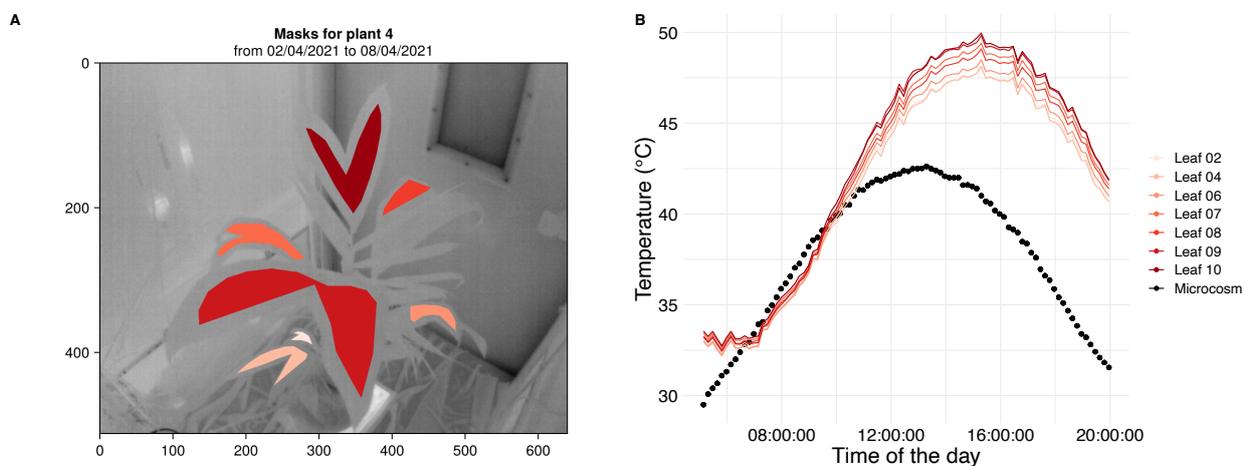

maximum, minimum and standard deviation of the temperatures within each mask were computed.

*Figure 8: A) Masks of leaf area for estimating leaf temperature. The mask is located at the centre of the leaf to avoid capturing image pixels that may not consistently represent the leaf due to internal chamber wind. The colours represent the different masks of the monitored leaves. B) Temperatures of leaves and the air temperature (black points) over a day.*

## Plant-level CO$_2$ and H$_2$O gas exchanges

The net flux of $CO_2$ was calculated from the inlet and outlet fluxes following (Eq1):

$$N = D(C_{in} - C_{out}) \qquad \text{(Eq1)}$$

Where N is the net flux of $CO_2$, in µmol s$^{-1}$, D is the flow rate of air at the inlet of the chamber, in µmol s$^{-1}$, $C_{in}$ and $C_{out}$ are the mixing ratio of $CO_2$ corrected for dilution by water vapour, respectively at the inlet and outlet of the chamber, in µmol mol$^{-1}$.

Because the pot of the plant was sealed, any variation in the weight measured from the scale could be attributed to plant transpiration. Large increases in pot weight were used as indicators of irrigation. Transpiration was then estimated using two different methods. The first method measured the difference in pot weight between the start and end of a specified time interval. The second, known as the regression method, determined transpiration by calculating the slope of a linear regression fitted to all weight measurements within that interval.

Transpiration and $CO_2$ fluxes showed contrasting variation during the day depending on the climate scenarios (Figure 9). The highest transpirations were recorded under the reference scenario and the 'dry hot' scenario. Interestingly, the 'hot' scenario showed the lowest level of transpiration and $CO_2$ assimilation. The integration over a day of the flux allowed to better compare the balance of assimilated carbon dioxide and transpired water depending on scenarios and plants (Figure 10). Results showed consistent behaviour among plants, with the highest value of assimilation under 600 ppm of [$CO_2$] and the lowest under the 'hot' scenario. The water use efficiency, calculated as the ratio of assimilated carbon dioxide over the transpired water, showed to be the highest under 800ppm and the lowest under the 'dry hot' scenario.

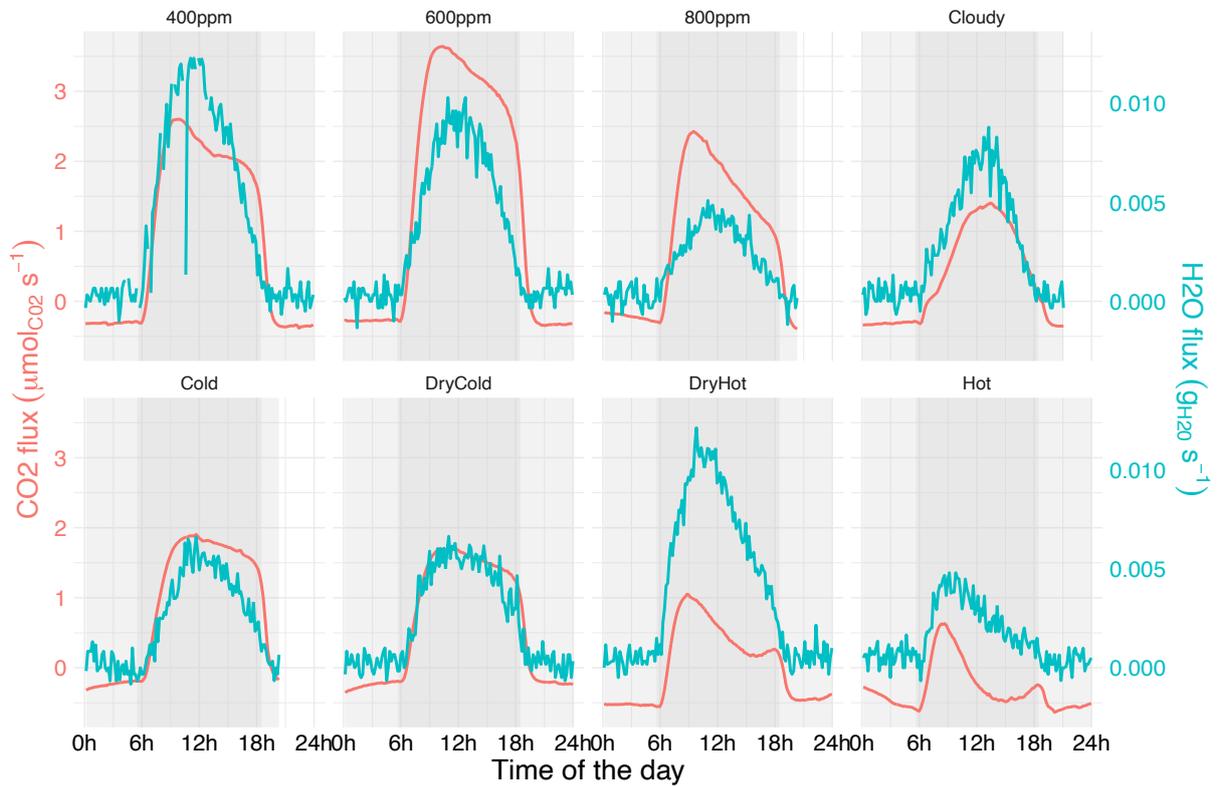

*Figure 9: Measured CO2 and H2O fluxes over a day for a single plant depending on the climate scenario. Each facet represents the fluxes for one scenario.*

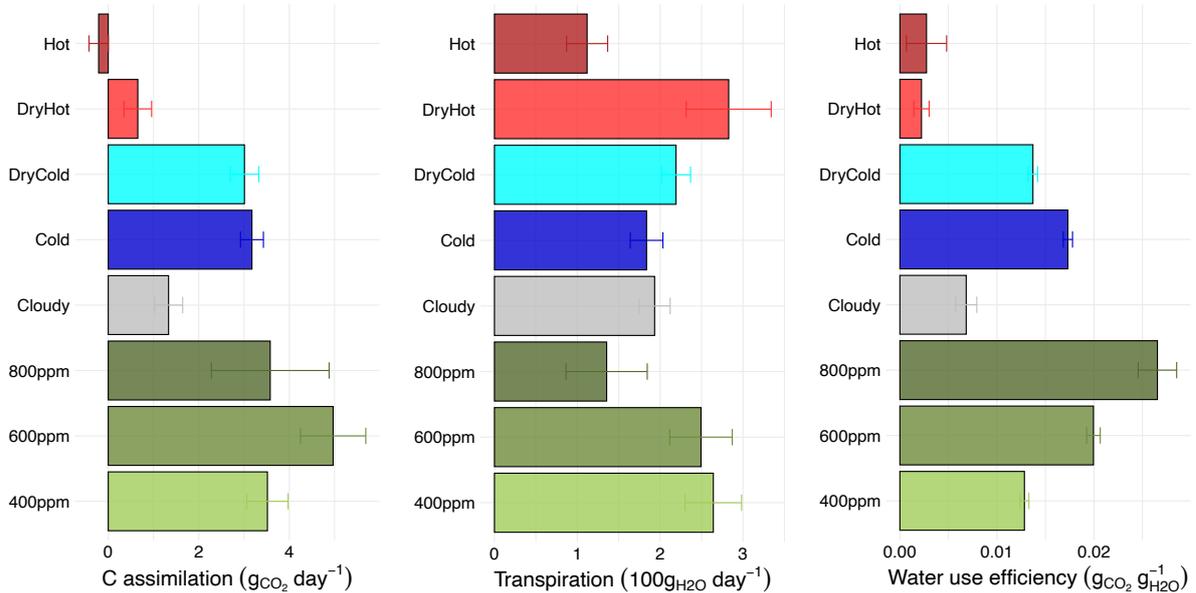

*Figure 10: Daily measurements of $CO_2$ assimilation, transpiration and water use efficiency depending on climate scenario. Bars represent the average value and error bars the standard deviation calculated from the four plants.*

The output data presented here demonstrated a gradient in physiological responses to the environmental conditions, providing a consistent foundation for evaluating simulation results generated by reproducing a digital twin oft he experiment through a Functional Structural Plant Model (FSPM).

## Impact of plant light environment on leaf-level gas exchange measurements

Leaf gas exchange measurements were conducted on plants located in the microcosm, where there were significant variations in the incoming light to the plants. Although the photosynthetic photon flux density (PPFD) received by the plants at mid-height in the chamber was low, small fluctuations in the light source affected the measured assimilation at the leaf scale (Figure 11). Assimilation varied both when the light in the Licor head chamber was saturated (WalzClose) and when it followed the incoming light of the microcosm (WalzOpen). Assimilation dropped sharply in response to changes in microcosm light, then increased again with a slight delay after light conditions improved. However, these changes in assimilation were less pronounced under WalzClose than under WalzOpen. These results highlight the sensitivity of leaf-scale measurements to the overall light environment of the plant and emphasize the importance of maintaining consistent and equivalent ambient light conditions at the plant scale when performing leaf-scale assimilation measurements to calibrate photosynthesis models or compare physiological performances.

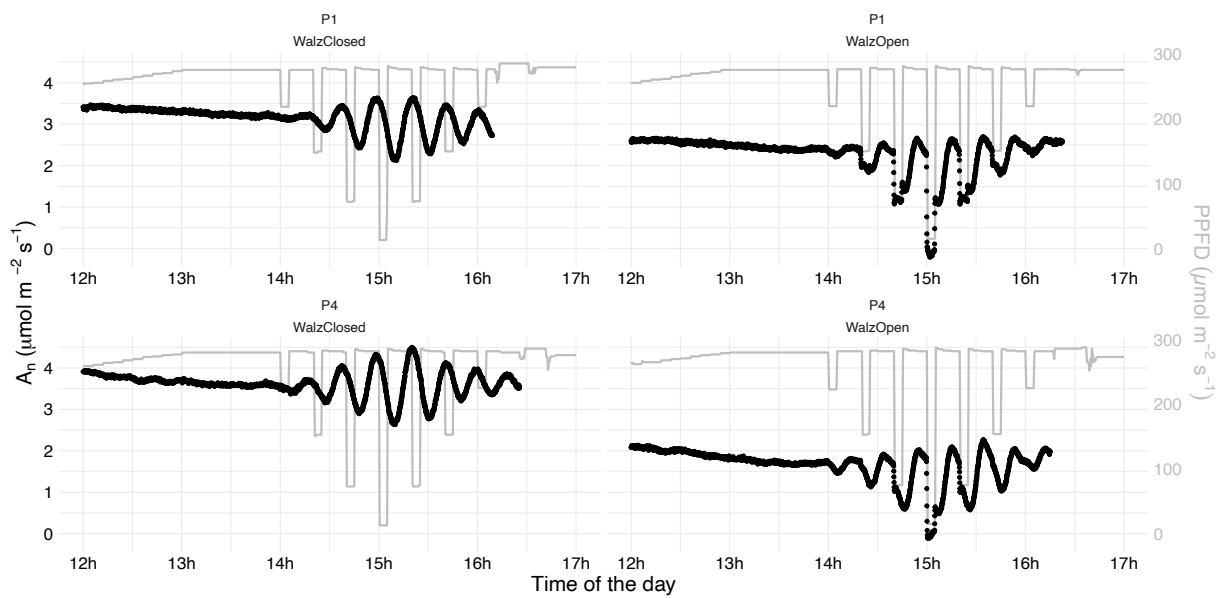

Figure 11: Net carbon assimilation (An) over time for two plants (P1 and P4) under saturating light conditions (WalzClosed, 1500 µmol m$^{-2}$ s$^{-1}$) or ambient microcosm light (WalzOpen) transmitted to the measured leaf in the Licor head chamber. The grey lines represent the ambient light measured by the PAR sensor in the microcosm (right axis).

## Data and code availability

The raw data and scripts used to generate the final database are detailed and accessible on Zenodo (https://doi.org/10.5281/zenodo.12705929), the code is also accessible via a Github repository (https://github.com/PalmStudio/Biophysics_database_palm), and we also provide a companion website (https://palmstudio.github.io/Biophysics_database_palm) showing how computations were made and the main results. The code to trigger the FLIR camera and for logging the precision scale data is also available on dedicated Zenodo repositories (https://doi.org/10.5281/zenodo.14862498 & https://doi.org/10.5281/zenodo.14862494).